\documentclass[twocolumn]{elsart}
\usepackage{epsfig}

\begin{document}

\begin{frontmatter}



\title{Single Crystal Growth of Skutterudite CoP$_3$ under High Pressure}


\author[aist]{C. H. Lee},
\author[aist]{H. Kito},
\author[aist]{H. Ihara},
\author[muroran]{K. Akita},
\author[muroran]{N. Yanase},
\author[muroran]{C. Sekine} and
\author[muroran]{I. Shirotani}
\address[aist]{AIST, 1-1-1 Umezono, Tsukuba, Ibaraki 305-8568, Japan}
\address[muroran]{Muroran Institute of Technology, 27-1 Mizumoto, Muroran
050-8585, Japan}

\begin{abstract}
A new method to grow single crystals of skutterudite compounds is 
examined.  Using a wedge-type, cubic-anvil, high-pressure apparatus, 
single crystals of CoP$_3$ were grown from stoichiometric melts 
under a pressure of 3.5 GPa.  Powder x-ray diffraction and electron 
probe microanalysis measurements indicate that the as-grown boules 
are a single phase of CoP$_3$.  The results suggest that CoP$_3$ is a 
congruent melting compound under high pressure.
\end{abstract}

\end{frontmatter}

\section{Introduction}
Filled skutterudite compounds, RM$_4$X$_{12}$ (R = rare earth, M = Fe, Ru 
or Os; X = P, As or Sb), are of interest to the scientific community 
because of their potential as thermoelectric materials and the 
instability of their $f$ electrons.  It is believed that the $f$ electron 
instability contributes to a wide variety of phenomena in the 
skutterudites, such as superconductivity, a metal-insulator 
transition \cite{sekine97,lee2001,lee99} and magnetic ordering.

To clarify the nature of the $f$ electron instability, as well as 
to develop the thermoelectric performance of skutterudites, 
studies based on single crystals are essential.  In particular, 
to improve the thermoelectric properties, it is important to 
clarify the origin of their superior thermoelectric performance.  
Filled skutterudites exhibit a remarkably low thermal conductivity, 
which enhances their performance.  Great efforts, thus, have been 
made to clarify the origin of this low thermal conductivity.  One 
proposed model that could explain the phenomenon is the so-called 
rattling effect.  It is conjectured that phonons are scattered 
effectively by the free vibration of rare earth atoms in large 
lattice cages \cite{caillat,sales}.  To examine the rattling effect, phonon 
behavior should be studied by neutron scattering, which requires 
large single crystals.

To date, a number of groups have grown single crystals of various 
skutterudites.  According to their reports, skutterudites are 
incongruent melting compounds.  In the growth process, Sb or Sn 
were typically used as flux, and samples were sealed in evacuated 
silica tubes.  The volume of grown binary skutterudites, such as 
CoSb$_3$ was typically 0.6 cm$^3$ \cite{caillat96}.  For ternary filled skutterudites, 
on the other hand, a typical sample size was about $1 \times 1 \times 1$ mm$^3$ 
\cite{jeitschko,torikachvili}, considerably smaller than that of binary skutterudites.

In the growth of skutterudite single crystals by the flux method, 
however, several limitations have been identified.  First, only a 
few skutterudite compounds have been able to be crystallized.  
This is a disadvantage in a comprehensive study of physical properties, 
where single crystals of a number of different skutterudites are 
necessary.  Second, the size of the grown crystals of ternary filled 
skutterudites is limited.  Larger single crystals are necessary in 
order to study fundamental properties by some methods, for example, 
by neutron scattering measurements.  To solve these problems, the 
development of an original approach to the growth of skutterudite 
crystals may be beneficial.
\begin{figure}
\includegraphics[width=\columnwidth]{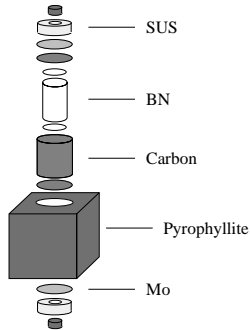}
\caption{\label{fig:sample-cell}A schematic illustration of the sample 
cell assembly used for crystal growth.}
\end{figure}

In this study, we analyze the growth of single crystals of skutterudites 
performed under high pressure using a wedge-type, cubic-anvil, 
high-pressure apparatus.  This technique was previously applied to enhance 
the growth of black phosphorus single crystals \cite{shirotani}.  Using this method, 
single crystals of black phosphorus of dimension $4 \times 2 \times 0.2$ mm$^3$ were 
successfully grown under a pressure of 2.3 GPa.  In this paper, we report 
that this technique is also useful for growing single crystals of skutterudites.

\section{Experimental Details}
CoP$_3$ single crystals were grown using a wedge-type, cubic-anvil, 
high-pressure apparatus (CAP-07, RIKEN).  Fig. 1 shows a schematic 
representation of the sample cell assembly used in the experiment.  
The sample container is a cube composed of pyrophyllite with sides of 
length 15.3 mm.  A carbon heater tube and a BN (boron nitride) crucible 
were inserted into the container.  The size of the BN crucible was 6 mm 
in length and 4.5 mm in diameter.  To estimate the temperature in the 
crucible during crystal growth, the relationship between the applied 
electrical power and the temperature was predetermined without 
encasing the samples.  Temperatures were measured up to $\sim1100^{\circ}$C 
at the center of the carbon tube using Chromel-Alumel thermocouples 
at ambient pressure without employing any emf correction for the pressure 
effect.  Temperatures above $1100^{\circ}$C were estimated by extrapolating the 
obtained temperature calibration curve.

The raw materials of Co and P of 99.9
crucibles after mixing them in a stoichiometric ratio.  The filled crucibles 
were then immediately placed in the container and compressed to 3.5 GPa 
at room temperature.  The samples were then heated to $\sim1100^{\circ}$C and sintered 
for 1 hour to pre-synthesize CoP$_3$ polycrystals.  Subsequently, the samples 
were heated to $\sim1500^{\circ}$C and maintained at that temperature for 1 hour.  
Finally, the samples were cooled to $1000^{\circ}$C over a period of 5 hours, 
followed by quenching to room temperature before pressure release.

The grown crystals were characterized by the following methods.  Phase 
identification of the grown crystals was performed by powder x-ray 
diffraction (RINT-1000, RIGAKU) at room temperature using Cu Ka radiation.  
The composition of the grown crystals, as well as the distribution of the 
Co and P atoms, was examined by a JEOL JSM-6301F scanning electron 
microscopeÐx-ray energy dispersion spectrometer (SEM-EDS) with an 
accelerating voltage of 20 kV.  A Li doped Si crystal was used as the 
SEM-EDS detector with an ultra-thin polymer window (CDU-Super-UTW, EDAX), 
that could detect elements ranging from .Be to U.  To estimate the 
concentration of Co and P atoms, characteristic x-rays of Co Ka and P Ka were used.  
A mapping analysis of the Co and P atoms was conducted with a scanning 
time of 10 ms/point and a measurement point density of 2500 points/mm$^2$ 
over the entire boule cross-section.  For the SEM-EDS measurements, the 
as-grown crystals were mechanically polished to mirror-like-surfaces.  
Back-reflection Laue x-ray measurements were performed with a collimator 
1 mm in diameter to confirm that the grown crystals were single crystals.
\begin{figure}
\includegraphics[width=\columnwidth]{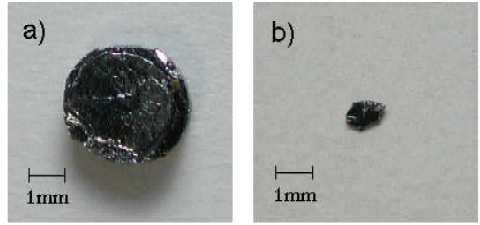}
\caption{\label{fig:photo}a) An as-grown boule of CoP$_3$.  b) A single 
crystal of CoP$_3$ separated from the as-grown boule.}
\end{figure}

\section{Results}
Fig. 2(a) shows a cross section of the as-grown boule of CoP$_3$.  The shape 
of the boule is cylindrical, about 3.4 mm in diameter, with a number of 
cracks.  The as-grown boule split quite easily along the radial direction to 
generate specimens $\sim1$ mm in diameter and $\sim0.3$ mm in length (Fig. 2(b)).  
Fig. 3 shows a back-reflection Laue x-ray photograph of the specimen displayed 
in Fig. 2(b).  In the photograph, 2mm symmetry is observed, indicating a (100) 
crystallographic plane.  This result suggests that the specimen is a single crystal.

Powder x-ray diffraction measurements were carried out using powder ground 
from the as-grown boule (Fig. 4).  The observed diffraction lines are all 
\begin{figure}
\includegraphics[width=\columnwidth]{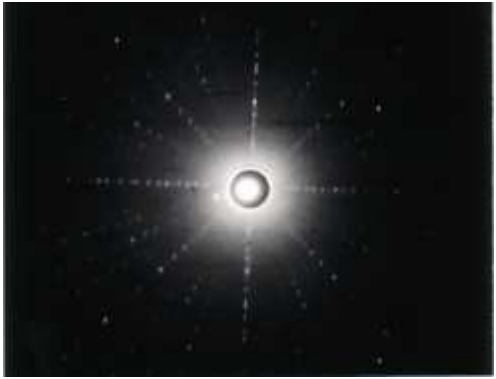}
\caption{\label{fig:laue}A back-reflection Laue x-ray photograph 
of a grown CoP$_3$ crystal.}
\end{figure}
indexable using the skutterudite structure (space group Im$\bar{3}$).  No impurity phases 
are observed, suggesting that the entire as-grown boule is a single phase of 
CoP$_3$.  The lattice constant determined by a least squares fit to the data was 
a = 7.7067 , which is close to the literature value \cite{rundqvist}.  This agreement 
suggests that the grown crystals are of high purity.

Fig. 5 shows the distribution of Co and P atoms within the as-grown boule 
measured by SEM-EDS.  The presence of the darker regions in the map is due 
to cracks, presumably formed during the cooling process.  In the other regions, 
a homogeneous distribution of Co and P atoms is observed.  We checked carefully 
that there were no additional elements present, especially around the cracks.  
The ratio of the Co to the P atoms was determined by averaging several measurements 
and was found to be Co:P = 0.97:3.00, which is consistent with a stoichiometric 
composition within instrumental error.  These results suggest that the crystals 
are a high purity, single phase of CoP$_3$.
\begin{figure}
\includegraphics[width=\columnwidth]{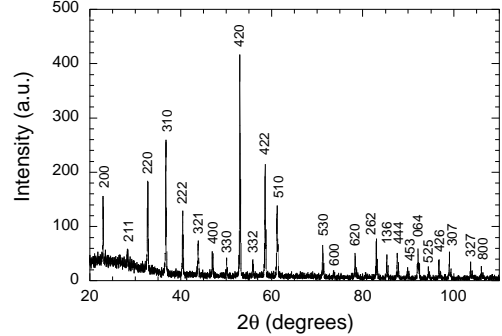}
\caption{\label{fig:diff}An x-ray diffraction pattern for a CoP$_3$ 
as-grown boule ground into powder.}
\end{figure}

\section{Discussion}
In this study, single crystals of CoP$_3$ were grown from a stoichiometric melt 
under high pressure using a wedge-type, cubic-anvil, high-pressure apparatus.  
The entire as-grown boule was characterized as a single phase of CoP$_3$ by 
powder x-ray diffraction and SEM-EDS measurements.  These results suggest 
that CoP$_3$ is a congruent melting compound under high pressure.  Around 
ambient pressures, however, it is still unclear whether CoP$_3$ is a congruent 
melting compound or not, since the phase diagram of Co-P is incomplete \cite{ishida-a}.  
Note that other skutterudites whose phase diagrams are well known, for 
example CoAs$_3$ and CoSb$_3$, are all incongruent melting compounds around ambient 
pressures \cite{ishida-b,ishida-c}.  It is, thus, natural to assume that CoP$_3$ is also an incongruent 
melting compound around ambient pressures in contrast to the results under high 
pressure.  Most likely, CoP$_3$ transforms from an incongruent to a congruent 
melting compound by the application of high pressures.  If this presumption is 
true, other skutterudites may also melt congruently under high pressure, which 
would be a great advantage for growing single crystals.
\begin{figure}
\includegraphics[width=\columnwidth]{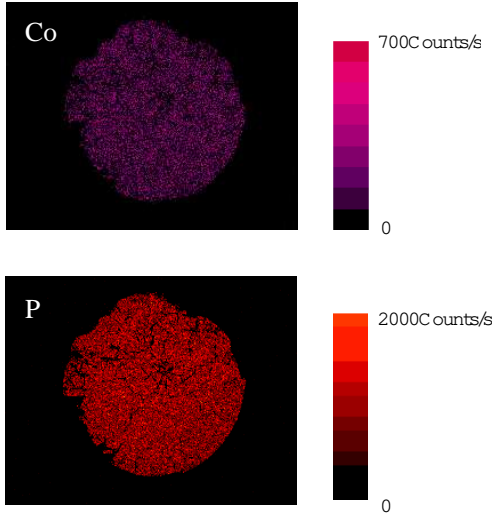}
\caption{\label{fig:mapping}Co and P distribution in a cross section of an 
as-grown boule of CoP$_3$ measured by SEM-EDS.}
\end{figure}

The size of the CoP$_3$ single crystals produced in this study is typically $\sim$ 1 
mm in diameter and $\sim0.3$ mm in length.  To enlarge the size of single crystals, 
the cracks in the as-grown boules need to be eliminated.  One possible solution 
to prevent cracking is to use a slower cooling rate.  Another solution could be to 
apply a more homogeneous pressure by using a smaller crucible.  Crystal growth 
from a single seed is also important for obtaining large single crystals.  For this 
to take place, the temperature distribution in the furnace must be further optimized.

\section{Conclusion}
Single crystals of CoP$_3$ were grown from a stoichiometric melt 
under a pressure of 3.5 GPa using a wedge-type, cubic-anvil, 
high-pressure apparatus.  The size of the grown single crystals is 
typically $\sim1$ mm in diameter and $\sim0.3$ mm in length.  The as-grown 
boules were characterized by powder x-ray diffraction and SEM-EDS 
measurements and were determined to be a single phase of CoP$_3$.  
These results suggest that CoP$_3$ melts congruently under high pressure.

\section{Acknowledgements}
The authors thank A. Iyo for valuable technical suggestions.  We also thank 
A. Negishi and A. Yamamoto for help in performing SEM-EDS measurements.  
This work was supported by a Grant from the Ministry of Economy, Trade and 
Industry of Japan.



\end{document}